\documentclass[12pt,a4paper]{PoS}

\title{Large statistics study of the topological charge distribution in the SU(3) gauge theory}

\ShortTitle{Large statistics study of QCD topological charge distribution}

\author{Leonardo~Giusti\\
   CERN, Department of Physics, TH Division, CH-1211 Geneva 23, Switzerland\\
   E-mail: \email{Leonardo.Giusti@cern.ch}}
   
\author{\speaker{Silvano~Petrarca} \\
   Dip. di Fisica, Universit\`a di Roma ``La Sapienza'', P.le 
                 A. Moro 2, I-00185 Rome, Italy\\
   INFN, Sezione di Roma, P.le A. Moro 2, I-00185 Rome, Italy\\	 
   E-mail: \email{Silvano.Petrarca@roma1.infn.it}}

\author{Bruno~Taglienti\\
   INFN, Sezione di Roma, P.le A. Moro 2, I-00185 Rome, Italy\\
   E-mail: \email{Bruno.Taglienti@roma1.infn.it}}

\abstract{ We present preliminary results for a high statistics study of 
 the topological charge distribution in the $SU(3)$ Yang-Mills theory
 obtained by using the definition of the charge suggested
 by Neuberger fermions. We find statistical evidence
 for deviations from a gaussian distribution.
 The large statistics required
 has been obtained by using PCs of the INFN-GRID.
}

\FullConference{XXIVth International Symposium on Lattice Field Theory\\
       July 23-28, 2006\\
       Tucson, Arizona, USA}

\begin{document}

\section{Introduction}
We perform a high statistics study of the topological charge distribution in the $SU(3)$ Yang-Mills
theory adopting the definition suggested by Neuberger fermions as it is discussed in
a series of recent studies~\cite{Neu1}-\cite{Lus1}. The numerical studies have been initiated
 in ref.\cite{GLWW03,DP} and completed in 
 ref.~\cite{DGP04} where a systematic study at different volumes and 
values of the lattice spacing has been performed in order to obtain a precise and reliable
determination of the topological susceptibility. Their distributions are obtained by a 
population between 1500 and 3000 topological charge for each volume. 
This already
impressive number of configurations allowed them to measure the variance of the 
distribution which is given
 $\langle Q^2\rangle$ and the topological susceptibility at a $\sim 5 \% $
  level but it was not enough to show 
 non-gaussianity. Moreover in~\cite{DGP04} it has been numerically confirmed that a physical volume larger than $(1 {\rm fm})^4$ guarantees that finite-volume effects are negligible at their
statistical precision.

The aim of this study is to look for non-gaussianity
in the topological charge distribution of the $SU(3)$ Yang-Mills
theory. 
Here we study  three lattices at the same physical volume $\sim (1.12 {\rm fm})^4$ with about ten time more statistics than in ref.~\cite{DGP04} in order to  emphasize  the deviations from the gaussian distribution. We stress that in order
 to search for such very small subleading effects it is necessary to be sure that all the systematics of the calculation
 cannot either simulate or hide the effect and therefore the only reliable theoretical framework
 is the one provided by the topological charge  definition suggested from Ginsparg-Wilson fermions. 
 
 This challenging Monte Carlo calculation has been 
 recently made possible by a few important improvements in the algorithmic  sector which guarantee the reliability and the feasibility of high statistics \cite{GHLW02}. Moreover algorithms for zero mode counting with no contamination from quasi zero modes, optimized to run fast on a single
 processor, are now available. The great computer effort has been performed in the frame of the INFN GRID project which allowed us to use the computer resources shared in the scientific italian network provided by INFN along this year. 

 \section{Theoretical procedure}
In the following we use the standard plaquette action of the $SU(3)$ gauge field. The massless lattice Dirac operator
$D$ satisfies the Ginsparg-Wilson relation
$$\gamma_5 D + D \gamma_5= \bar{a} D \gamma_5 D \; .$$

In this frame the
topological charge density can be defined
 \begin{equation}
 q(x)= - {\bar{a} \over 2}Tr[\gamma_5 D(x,x)]
  \label{eq:chaden}
   \end{equation}
where  ${\bar a}=a/(1+s)$,  $a$ is the lattice spacing and
the shift  parameter $s$  has been fixed  in our calculation at the value $s=0.4$.
The topological charge is obtained from the lattice by computing for each
configuration   the difference between the numbers of
zero modes with positive and negative chiralities that is the index $\nu$ of the Dirac operator: 
 $$\nu =n_+ - n_- \; ,$$
 which is directly related to the topological charge $Q$:
 \begin{displaymath}
\nu=Q=a^4 \sum_x q(x) 
\end{displaymath}
 
As it is stressed in ref.\cite{GLWW03}  the asymptotic distribution of the topological charge
follows a gaussian form centered in zero, with the only parameter $\langle Q^2 \rangle$ as
variance :
  \begin{equation}
P_Q ={1\over {\sqrt {2 \pi \langle Q^2 \rangle}}} \exp{({- {Q^2} \over {2 
\langle Q^2 \rangle}})}   \; .    
 \label{eq:leasy}
 \end{equation}
 Corrections to this formula are expected to be suppressed by terms of $1/{N_c}^2$ and $1/V$
where $V$ is the lattice volume
and $N_c$ is the  number of colours of the theory.

\section{Our project}

In the Table \ref{tab:lattici} it is shown the summary of the features of our three runs altogether with the information about the occupancy of the Grid network. The number of
configuration are written in form of a product of the number of simultaneous runs sent to the Grid network and the total number of configurations planned for each run. The fourth column indicates the computer time taken for each configuration running on a single processor. 
The fifth column contains the number of configuration sequentially processed in a single run sent to the Grid network. In the case of the largest volume the desired number of configurations (30000)
was not achieved at the time of the Conference.
\vspace{0.5 cm}
\begin{table}[hpbt]
\begin{center}
\begin{tabular}{|c|c|c|c|c|c|c|}
\hline
 Lattice  & $ \beta$  &$r_0/a$&  Total  \# Confs   &  Time           &  \# Confs for each run   &  Ram (MB)  \\
\hline
$ 12^4  $ & $ 6    $   &  $5.368$&  $ 58 x 600$  & 1 h/conf      & $ 10       $  & $ 256  $   \\           
\hline
$ 14^4  $ & $ 6.0938 $ &  $6.263$& $ 120 x 250$ & 4 h/conf      & $ 4        $  & $ 512  $   \\
\hline
$ 18^4  $ & $ 6.2623 $&$8.052$ & $ 200 x 150$ &  12 h/conf      & $ 2        $  & $ 1100  $  \\
\hline
\end{tabular}
\end{center}
\caption{
Summary of our lattices, physical volume $\sim (1.12 fm)^4$,  $r_0=0.5$fm. The run of the $18^4$ lattice
was not completed at the time of the presentation.}
\label{tab:lattici}
\end{table}

We stress that all our runs are performed over a single processor using its local memory.
The parameters of the lattices are chosen  to  fix the physical volume at the value $\sim (1.12 fm)^4$ at 
three lattice spacing in order to estimate the size of the discretization effects.

\begin{figure}[h]
   \includegraphics[width=1\textwidth]{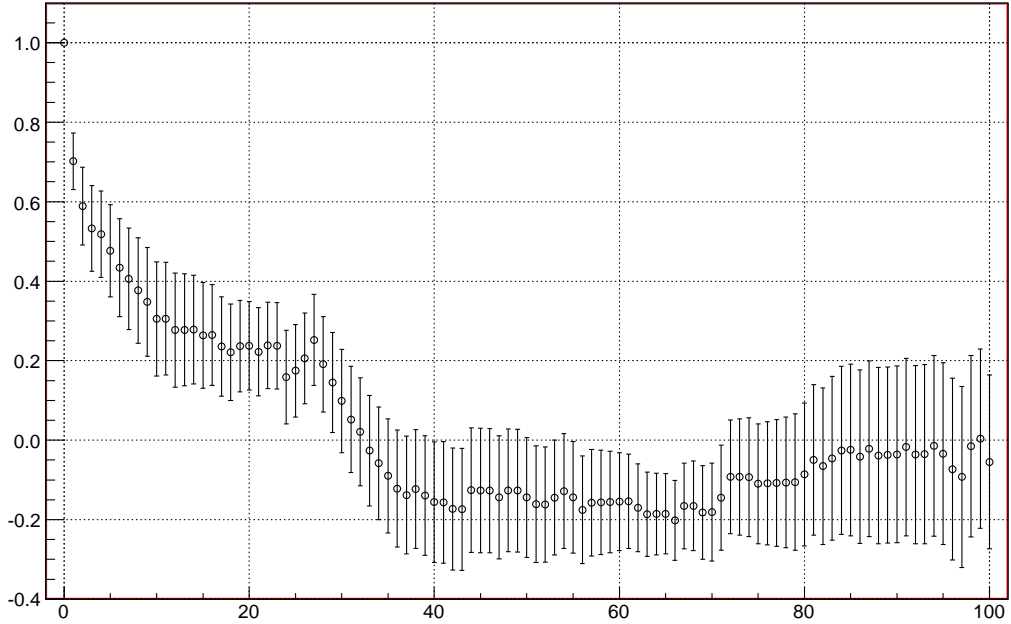}
   \caption{Autocorrelation of the topological charge as function of computer updates in the
   worst case of the largest volume $18^4$.}
   \label{fig:c18}
   \end{figure}

In order to check the statistical independence of the measurements we have first studied the
topological charge autocorrelation function. The worst case of the  $18^4$ lattice is shown in Fig.~\ref{fig:c18}.
 This figure shows a typical  autocorrelation time
of order of  about $40$ in units of Monte Carlo updates. 
Each update is composed of one heat-bath and 6, 7, 9 cycles  of over-relaxation 
of all the links for each volume.
We have chosen to separate the subsequent measurements by $500$, $1500$ and $2500$ updates for the volumes  $12^4$, $14^4$ and $18^4$ respectively.  This  large separation among measurements does not cost too much
in computer time and allows us to consider the measurements as statistically independent.

\section{Distributions at large statistics: preliminary data}
\begin{table}[hpbt]
\begin{center}
\begin{tabular}{|c|c|c|c|c|c|}
\cline{3-6}
\multicolumn{2}{c|}{}
 &\multicolumn{2}{|c|}{Data from ref.~\cite{DGP04}}
&\multicolumn{2}{|c|}{Our preliminary data} \\
\hline
$ \beta$  & $ L/a  $   &    \# Confs  &  $\langle Q^2 \rangle$     & $\langle Q^2\rangle$  & \# Confs \\
\hline
$ 6  $    & $ 12  $    & $ 2452$      &  $1.63(8) $              & $ 1.638 (13)     $  & $ 34800 $        \\           
\hline
$ 6.0938$ & $ 14 $     & $ 1405$      &   $1.54 (6) $            & $ 1.565 (13)     $  & $ 30000  $       \\
\hline
$ 6.2623$ & $ 18 $     &              &                          & $1.442 (18)$        & $ 13561  $       \\
\hline
\end{tabular}
\end{center}
\caption{Comparison with ref.~\cite{DGP04} for the first two lattices. The three physical sizes have
about the same value $L\sim 1.12$ fm. }
\label{tab:Comparison}
\end{table}
In the Table \ref{tab:Comparison} we show our { preliminary} data for the correlator $\langle Q^2 \rangle$ in comparison with the results of ref. \cite{DGP04}. The values are in complete agreement each other and moreover it is clear that the errors scale with the $1/\sqrt N$
law as it is expected due to their statistical nature. 

\begin{figure}[ht]
  \begin{tabular}{cc}
   \includegraphics[width=.5\textwidth]{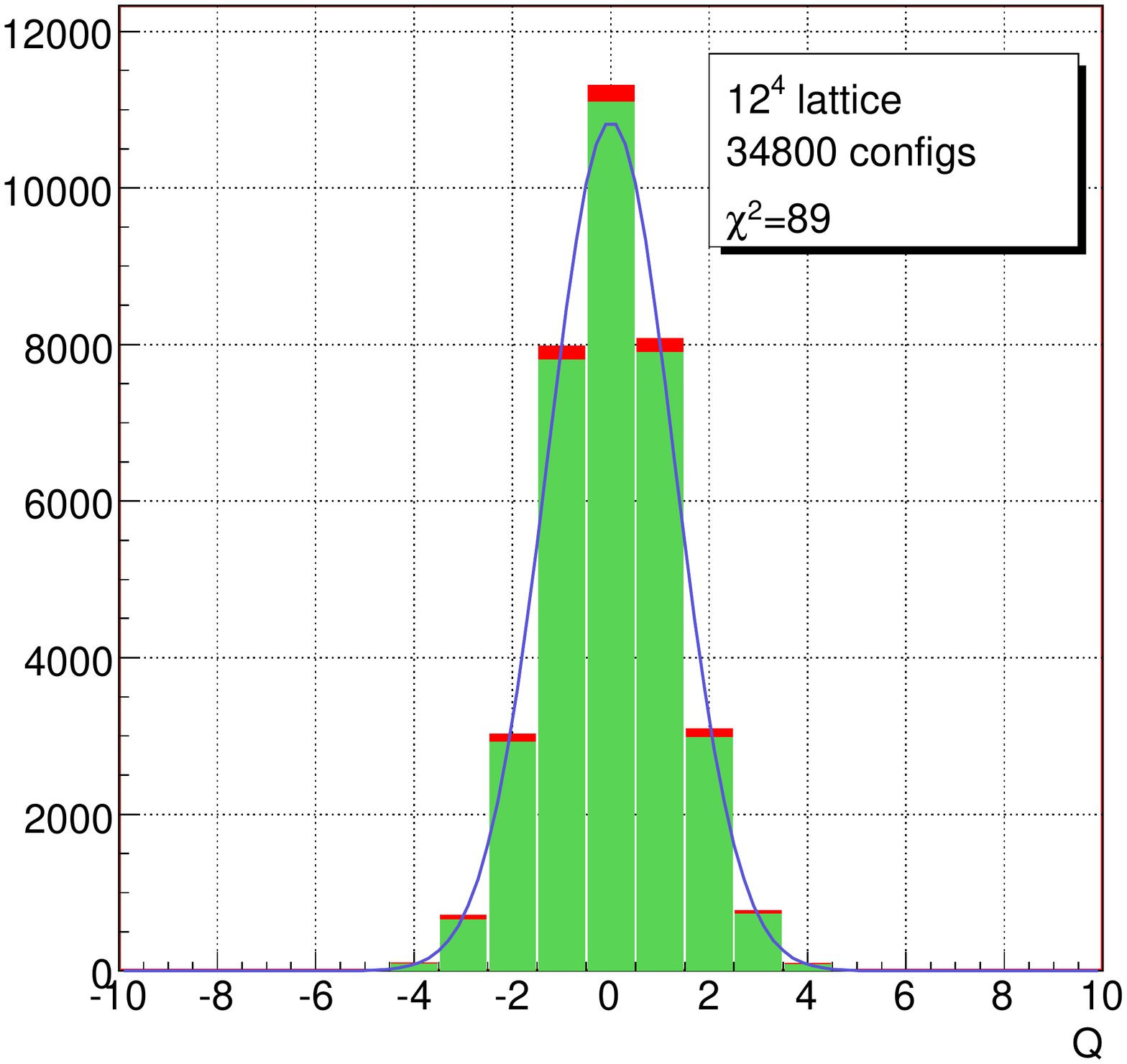} 
   \includegraphics[width=.5\textwidth]{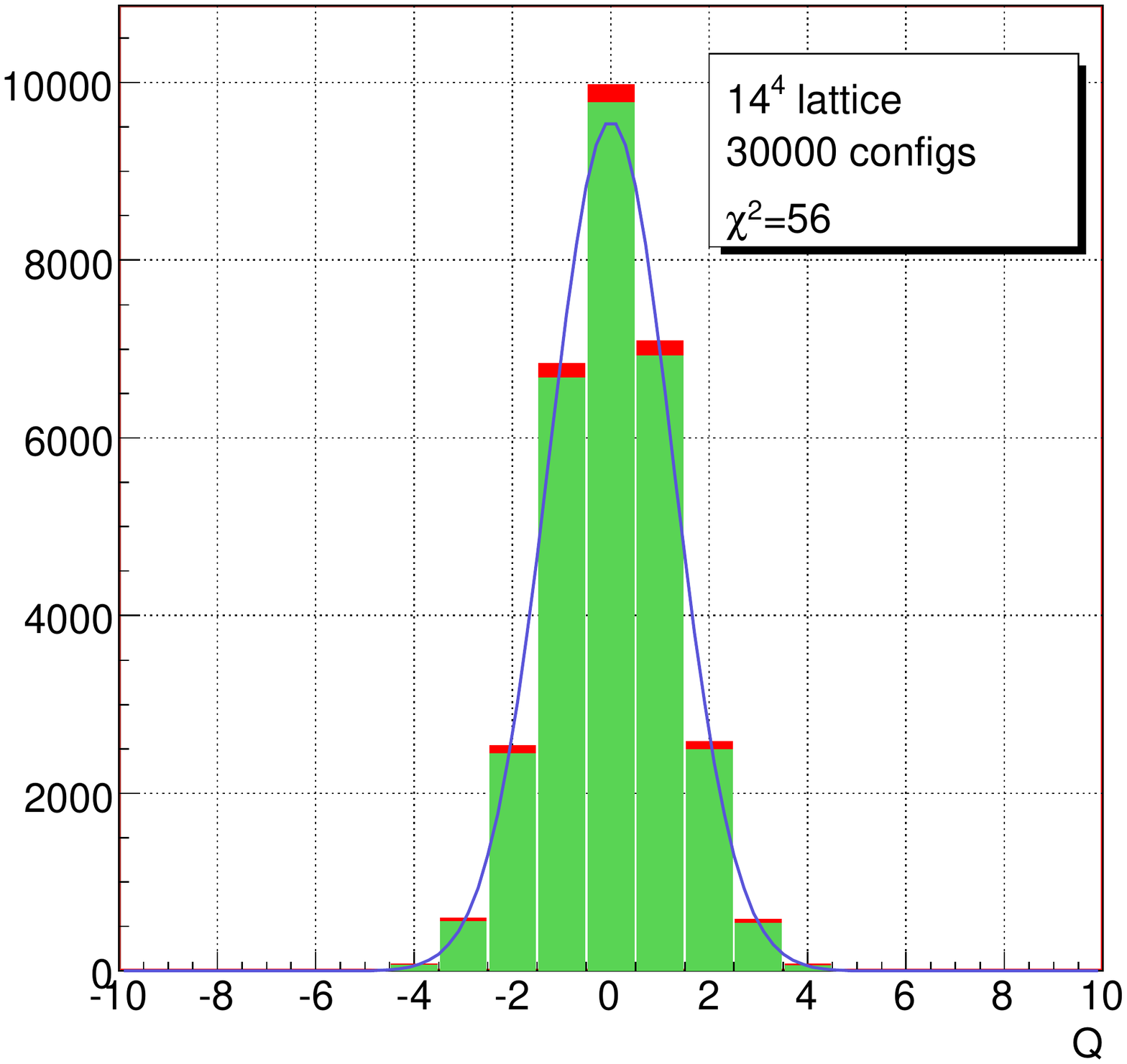}\\
\end{tabular}
   \caption{Histograms of the topological charge for $12^4$ and $14^4$  lattices. The border at the top indicates the statistical error of the column. In each picture the value of the
   $\chi^2$ is reported. }
    \label{fig:121418}
   \end{figure}

In order to give a graphic representation of our data, following~\cite{DGP04},  we show in Fig.~\ref{fig:121418} the histograms of the topological charge of the first two lattices for which the runs were completed at the time of the presentation.  
We underline that  the topological charge is a discrete variable and  therefore the gaussian form must not be confused with the normal distribution even if
the  mathematical expression is the same. Here plotting the histograms of the topological charge values in comparison with the continuos line of the gaussian, we want only to give  a graphical representation of the data and the comparison with the asymptotic formula must be done comparing the value at the center of the column with the level of the column. The border  at the top indicates the statistical error ($1$ sigma) of the column. At first glance it is worth to note that the data at $Q=0$ are sensitively higher than the theoretical gaussian value.

\begin{figure}[ht]
  \begin{tabular}{cc}
   \includegraphics[width=.5\textwidth]{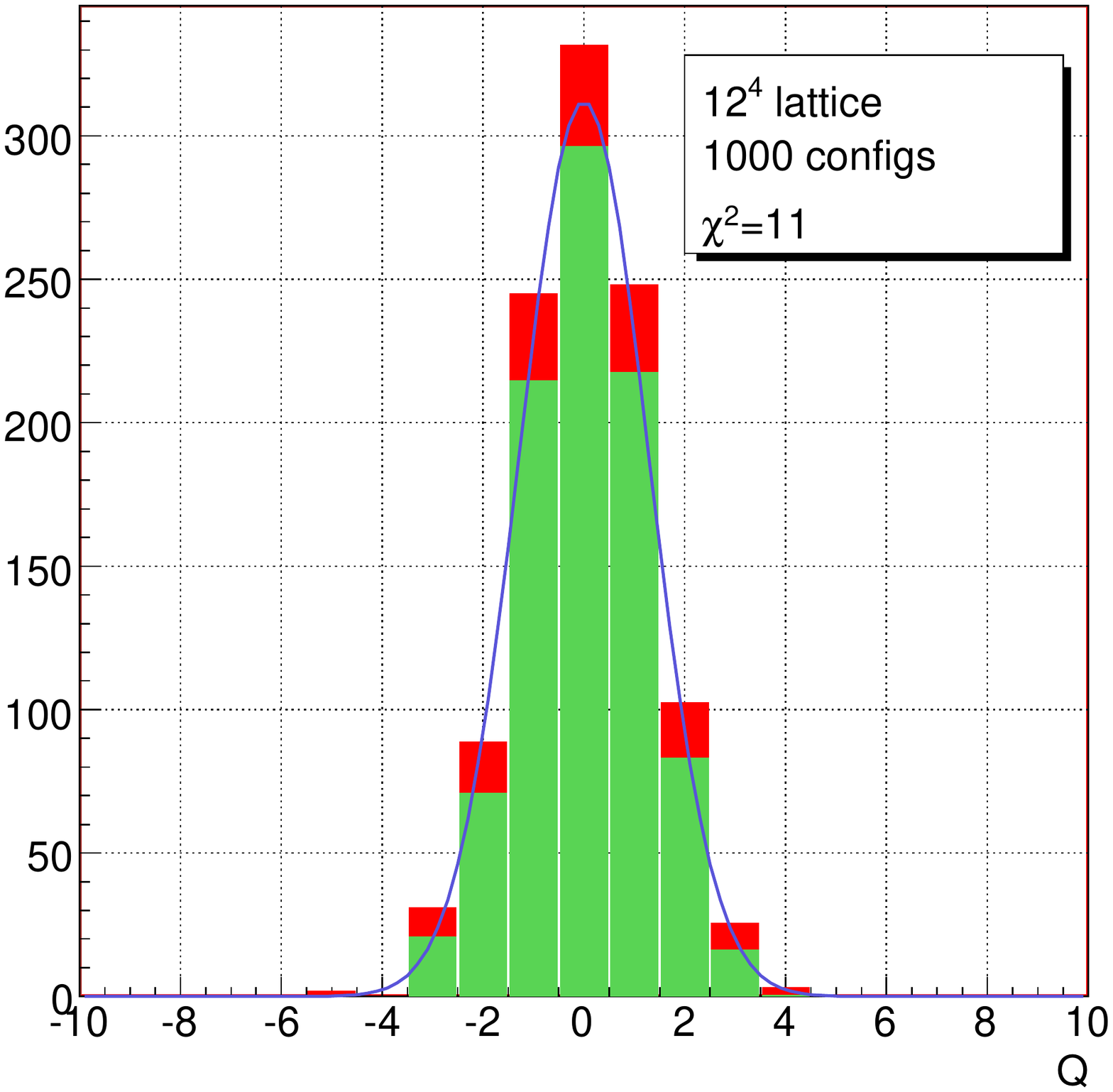}
   \includegraphics[width=.5\textwidth]{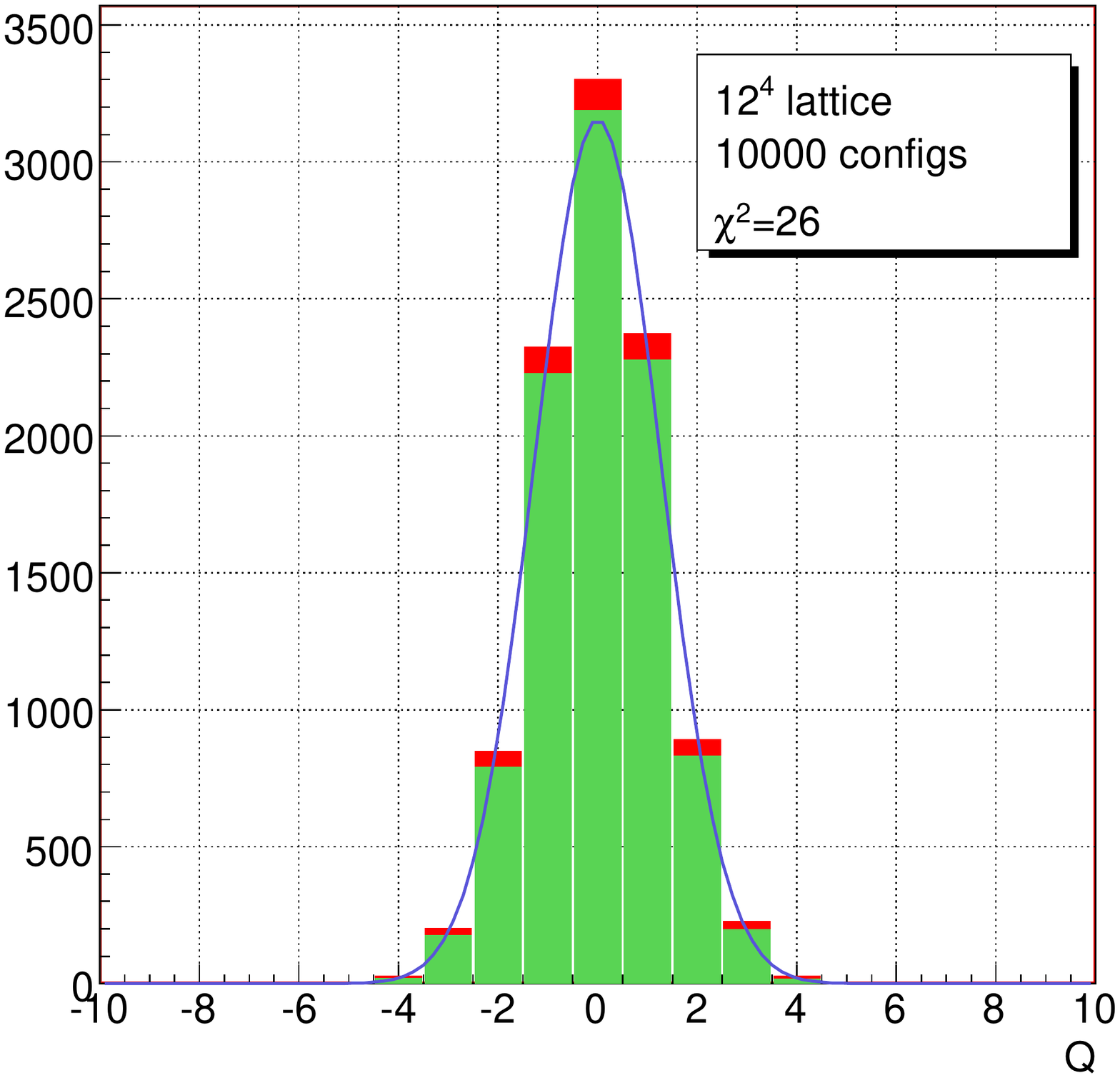}
   \end{tabular}
   \caption{Histograms with 1000 and 10000 configurations for the $12^4$ run. In the first  sample the data are statistically compatible with  a pure gaussian while for the second one the data 
    begins to show a deviation of more than 3 sigmas from a pure gaussian.}
   \label{fig:12medium}
   \end{figure}

To be quantitative we test the hypothesis that large statistic data are
compatible with the leading asymptotic distribution eq.(\ref{eq:leasy}) taking as
parameter the measured value of $\langle Q^2 \rangle$. We use the test of $\chi^2$, value that is reported in each of the pictures in Fig.~\ref{fig:121418}. 
The value of $\chi^2$ for  the data at large statistics is greater than the critical value value~\footnote{The critical value for a $\chi^2$ distribution with 12 degrees of freedom  is 14, 30, 52 for 1, 3, 5 standard deviations respectively.} corresponding to more than  five standard deviations therefore the data are not compatible with the pure gaussian distribution of the topological charge.\\
In order to study the behaviour as function of the number of collected topological charges
we show in Fig. \ref{fig:12medium} for the smallest volume $12^4$ the histograms at increasing statistics.
We plotted two histograms of 1000 and 10000 values. The $\chi^2$ test indicates that the first sample is compatible with a pure gaussian within one sigma while in the second sample the data show a larger $\chi^2$ value corresponding to a deviation of more than three sigmas and therefore the test-hypothesis should  be rejected.

The most interesting result we have to show is the value of the 
fourth cumulant of the topological charge distribution:
\begin{displaymath}
\kappa_4=\langle Q^4 \rangle - 3 \langle Q^2 \rangle^2\; .
\end{displaymath}
This quantity is zero for a pure gaussian distribution and the previous data from ref.~\cite{DGP04}
are consistent with zero. The values we found are definitely different from zero. We quote preliminarly:
$$  \quad  \kappa_4({\rm lattice}12^4)= 0.609 \pm 0.093 \quad
 {\rm and} \quad
\quad  \kappa_4({\rm lattice}14^4)= 0.442\pm 0.078$$
where the errors are  1  jacknife standard deviation.

\section{Outlook}

We plan to finish  the runs of the third volume in a few weeks reaching the prefixed value of 30000 configurations, then we will complete the checks in particular the dependence from the volume size, for which we plan to run a test volume at $16^4$, $\beta= 6$ with $L \sim 1.5$fm. 

Even though the analysis presented and the results are still preliminary they
already  shown that the topological charge distribution presents deviations from gaussianity.
This goal has been reached with a statistics of 30000 configurations. 

We warmly thank Giuseppe Andronico for the great organization of Theophys, the virtual organization of  INFN Grid Project for theoretical physics.
We thank Alessandro De Salvo and Marco Serra of the INFN sez.Rome for the continuos effort in 
helping us during the accomplishment of the project.
We thank L.~Del~Debbio for his interest in an early stage of  this calculation.

\end{document}